\documentclass[manuscript, screen]{acmart}

\usepackage{xcolor}
\usepackage{subcaption}
\definecolor{pastelBlue}{RGB}{200, 230, 240}
\definecolor{pastelGreen}{RGB}{200, 245, 200}
\definecolor{pastelYellow}{RGB}{255, 245, 200}
\definecolor{pastelPink}{RGB}{255, 210, 220}
\definecolor{pastelPurple}{RGB}{235, 220, 240}
\definecolor{pastelOrange}{RGB}{255, 225, 200}
\definecolor{pastelMint}{RGB}{210, 250, 225}
\definecolor{pastelGray}{RGB}{235, 235, 235}
\definecolor{pastelTeal}{RGB}{210, 245, 245}
\definecolor{pastelCoral}{RGB}{255, 200, 185}

\definecolor{darkBlue}{RGB}{30, 70, 120}
\definecolor{darkGreen}{RGB}{20, 100, 50}
\definecolor{darkYellow}{RGB}{180, 140, 20}
\definecolor{darkPink}{RGB}{170, 40, 80}
\definecolor{darkPurple}{RGB}{90, 40, 110}
\definecolor{darkOrange}{RGB}{180, 80, 20}
\definecolor{darkMint}{RGB}{0, 120, 90}
\definecolor{darkGray}{RGB}{80, 80, 80}
\definecolor{darkTeal}{RGB}{0, 100, 110}
\definecolor{darkCoral}{RGB}{190, 70, 50}

\definecolor{prs-color}{HTML}{9C7BB5}
\definecolor{d-color}{HTML}{7FA0D8}
\definecolor{a-b-color}{HTML}{6FAE9C}
\definecolor{c-color}{HTML}{9A9A9A}

\usepackage[table]{xcolor}
\definecolor{lightgray}{gray}{0.92}

\definecolor{Acolor}{RGB}{200, 220, 245}  % soft blue
\definecolor{Bcolor}{RGB}{200, 235, 230}  % blue-teal
\definecolor{Ccolor}{RGB}{210, 240, 215}  % cool mint green
\definecolor{Dcolor}{RGB}{225, 215, 245}  % cool lavender
\definecolor{L-color}{RGB}{245, 235, 200}  % warm sand / beige
\definecolor{M-color}{RGB}{245, 220, 200}  % warm peach
\definecolor{B-color}{RGB}{245, 205, 210}  % soft pastel red

\usepackage[most]{tcolorbox}

\newtcbox{\pastelbox}[2][]{on line,
  arc=3pt, boxrule=0pt, left=1pt, right=1pt, top=0.1pt, bottom=0.1pt,
  colback=#2, colframe=#2, #1}

\usepackage{pifont}
\usepackage{subcaption}

\newcommand{\yes}{\textcolor{green!60!black}{\large\ding{51}}}
\newcommand{\no}{\textcolor{red!70!black}{\large\ding{55}}}

\definecolor{figorange}{RGB}{243,138,97}
\definecolor{figblue}{RGB}{88,129,170}

\AtBeginDocument{%
  }

\setcopyright{none} % acmlicensed
\copyrightyear{2026}
\acmYear{2026}
\acmDOI{XXXXXXX.XXXXXXX}
\begin{document}

%%
%% The "title" command has an optional parameter,
%% allowing the author to define a "short title" to be used in page headers.
\title{\textsc{PromptHelper}: A Prompt Recommender System for Encouraging Creativity in AI Chatbot Interactions}

%%
%% The "author" command and its associated commands are used to define
%% the authors and their affiliations.
%% Of note is the shared affiliation of the first two authors, and the
%% "authornote" and "authornotemark" commands
%% used to denote shared contribution to the research.

\author{Jason Kim}
\affiliation{%
  \institution{Texas A\&M University}
  \city{College Station}
  \state{Texas}
  \country{U.S.A.}}
\email{jasenio@tamu.edu}

\author{Maria Teleki}
\affiliation{%
  \institution{Texas A\&M University}
  \city{College Station}
  \state{Texas}
  \country{U.S.A.}}
\email{mariateleki@tamu.edu}

\author{James Caverlee}
\affiliation{%
  \institution{Texas A\&M University}
  \city{College Station}
  \state{Texas}
  \country{U.S.A.}}
\email{caverlee@tamu.edu}

%%
%% By default, the full list of authors will be used in the page
%% headers. Often, this list is too long, and will overlap
%% other information printed in the page headers. This command allows
%% the author to define a more concise list
%% of authors' names for this purpose.
\renewcommand{\shortauthors}{Kim et al.}

%%
%% The abstract is a short summary of the work to be presented in the
%% article.
\begin{abstract}
Prompting is central to interaction with AI systems, yet many users struggle to explore alternative directions, articulate creative intent, or understand how variations in prompts shape model outputs. We introduce prompt recommender systems (PRS) as an interaction approach that supports exploration, suggesting contextually relevant follow-up prompts. We present \textsc{PromptHelper}, a PRS prototype integrated into an AI chatbot that surfaces semantically diverse prompt suggestions while users work on real writing tasks. We evaluate \textsc{PromptHelper} in a 2x2 fully within-subjects study ($N=32$) across creative and academic writing tasks. Results show that \textsc{PromptHelper} significantly increases users’ perceived exploration and expressiveness without increasing cognitive workload. Qualitative findings illustrate how prompt recommendations help users branch into new directions, overcome uncertainty about what to ask next, and better articulate their intent. We discuss implications for designing AI interfaces that scaffold exploratory interaction while preserving user agency, and release open-source resources to support research on prompt recommendation.
\end{abstract}

%%
%% The code below is generated by the tool at http://dl.acm.org/ccs.cfm.
%% Please copy and paste the code instead of the example below.
%%

%%
%% Keywords. The author(s) should pick words that accurately describe
%% the work being presented. Separate the keywords with commas.
%% A "teaser" image appears between the author and affiliation
%% information and the body of the document, and typically spans the
%% page.
\begin{teaserfigure} %% swap out with fig 3
    \centering
    \begin{subfigure}[t]{0.49\textwidth}
      \centering
      \includegraphics[width=\linewidth]{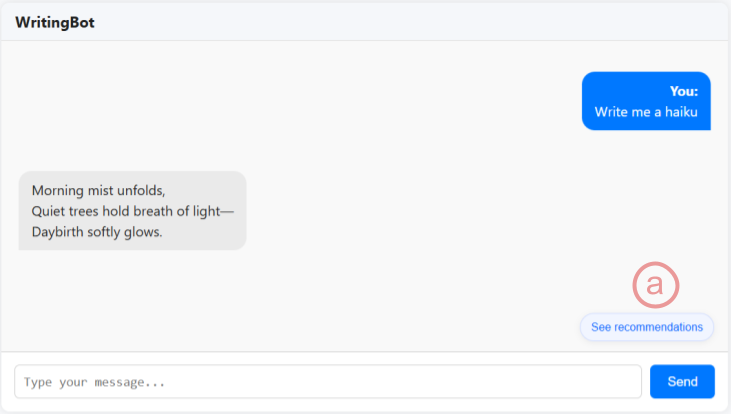}
      %\caption{First subfigure}
    \end{subfigure}
    \hfill
    \begin{subfigure}[t]{0.49\textwidth}
      \centering
      \includegraphics[width=\linewidth]{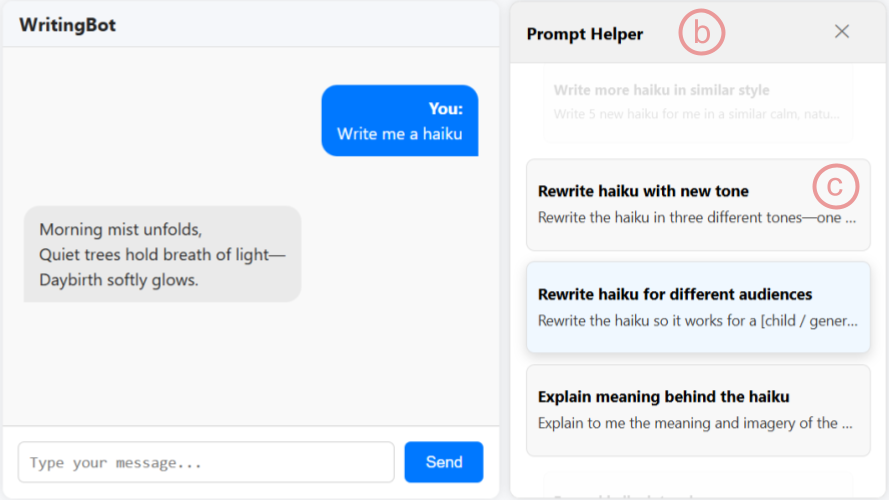}
      %\caption{Second subfigure}
    \end{subfigure}
    \caption{\textbf{Compared to a standard chatbot (left), \textsc{PromptHelper} (right) supports user interactions by encouraging creativity and diversity:} The user interacts with a baseline writing chatbot as normal. Clicking (a)  surfaces \textsc{PromptHelper}, (b), which displays contextually relevant follow-up prompt suggestions, based on the user’s last prompt and the model’s response. Suggested prompts, (c), highlight different creative directions (e.g., rewriting, audience adaptation, explanation) and can be copied and modified to support iterative exploration during writing.}
    \Description{-}
    \label{fig:teaser}
\end{teaserfigure}

\received{20 February 2007}
\received[revised]{12 March 2009}
\received[accepted]{5 June 2009}

%%
%% This command processes the author and affiliation and title
%% information and builds the first part of the formatted document.
\maketitle

\section{Introduction}
Generative AI systems increasingly enable end users to produce text, images, and other artifacts with minimal effort \cite{fui2023generative, wei2022emergent}. Despite this apparent accessibility, effective interaction with these systems continues to depend heavily on users’ ability to craft prompts \cite{sahoo2024systematic, schulhoff2024prompt, oppenlaender2025prompting, KNOTH2024100225}. Prior work shows that users often have a general sense of their intent, but struggle to translate that intent into precise language, articulate nuanced constraints, or systematically explore alternative directions \cite{subramonyam2024bridging, nguyen2024beginning}. As a result, interaction is frequently limited to a narrow subset of possibilities, constraining both output quality~\cite{barkley2024investigatingrolepromptingexternal} and creative exploration~\cite{sood2025conversationalinterfaceslimitcreativity}.
Most generative AI interfaces center interaction on a single prompt box. While this design appears flexible and lightweight, it places the burden of iteration, comparison, and exploration almost entirely on the user. Users must manually generate variations, recall prior attempts, and reason about how changes in phrasing affect model outputs, introducing substantial cognitive overhead \cite{subramonyam2024bridging, nguyen2024beginning}. At the same time, prompts themselves have emerged as rich interaction artifacts: they encode user goals, stylistic preferences, and strategies for steering model behavior.
\textbf{Treating prompts as recommender-eligible items therefore suggests new opportunities for supporting creative work with generative AI.}

In this paper, we propose prompt recommender systems (PRS) as an interaction approach for supporting user engagement with generative models. PRS provide users with alternative prompts, diverse starting points, and contextually relevant suggestions that help users broaden their search space and clarify creative intent~\cite{10.1145/3640794.3665881}. In contrast to approaches that automatically optimize or rewrite prompts on users’ behalf~\cite{wang2023reprompt, kong2024prewrite, li2024learning}, PRS are designed to complement existing chatbot interfaces by offering guidance while preserving user agency and control.

We instantiate this approach through \textsc{PromptHelper} (Figure~\ref{fig:teaser}), a PRS prototype integrated into an AI-assisted writing environment. \textsc{PromptHelper} presents context-aware follow-up prompts that reflect different creative directions, enabling users to compare alternatives and iteratively refine their prompts as their ideas evolve. We evaluate \textsc{PromptHelper} in a 2x2 fully within-subjects study ($N=32$) across creative and academic writing tasks, with the system enabled or disabled. The study combines quantitative measures of exploration, expressiveness, results-worth-effort, usability, and workload with qualitative feedback. Results show that \textsc{PromptHelper} increases perceived exploration and expressiveness without significantly affecting usability or mental workload. Participants report that the system helps them consider new ideas, overcome creative ruts, and better understand how to steer the model during writing.

We make the following contributions: \textbf{(1) A conceptualization of prompt recommender systems (PRS) (\S \ref{sec:prs})} as an interaction technique for supporting exploration, expressiveness, and creative workflows in generative AI tools; \textbf{(2) \textsc{PromptHelper}, a PRS prototype (\S \ref{sec:user-study})} that integrates prompt generation, diversification, and interactive refinement to support navigation of prompt space during writing tasks; \textbf{(3) A user study (N=32) (\S \ref{sec:user-study})}, that evaluates \textsc{PromptHelper} across two writing tasks: a creative writing task and an academic writing task. We report quantitative effects on exploration, expressiveness, results-worth-effort, workload, and usability; \textbf{(4) Open-sourced resources (\url{https://anonymous.4open.science/r/Prompt_Recommender-6B80})}, including system code, prompts, and analysis scripts to support research on prompt assistance and creative human-AI interaction.

% Image Examples from real tools
\begin{figure}[t]
\centering

% ---------- LEFT COLUMN ----------
\begin{subfigure}[t]{0.26\textwidth}
  \centering
  % {\large\bfseries Gemini \par}
  % \vspace{0.5em}
  \vspace{0.9em}

  \begin{subfigure}[t]{\linewidth}
    \centering
    \includegraphics[width=0.99\linewidth]{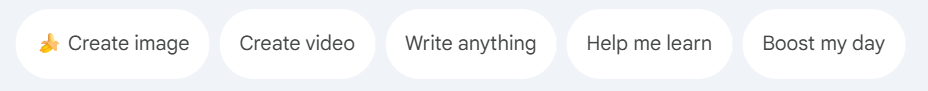}
    \caption{\textcolor{a-b-color}{Cold-Start Suggestions}, Gemini}
  \end{subfigure}

  \vspace{1.2em}

  \begin{subfigure}[t]{\linewidth}
    \centering
    \includegraphics[width=0.99\linewidth]{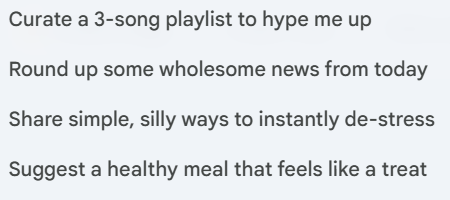}
    \caption{\textcolor{a-b-color}{Cold-Start Suggestions}, Gemini}
  \end{subfigure}
\end{subfigure}
\hfill
% ---------- Midele COLUMN ----------
\begin{subfigure}[t]{0.33\textwidth}
  \centering
  % {\large\bfseries ChatGPT \par}
  % \vspace{0.5em}
  \vspace{0.1em}

  \begin{subfigure}[t]{\linewidth}
    \centering
    \includegraphics[width=0.99\linewidth]{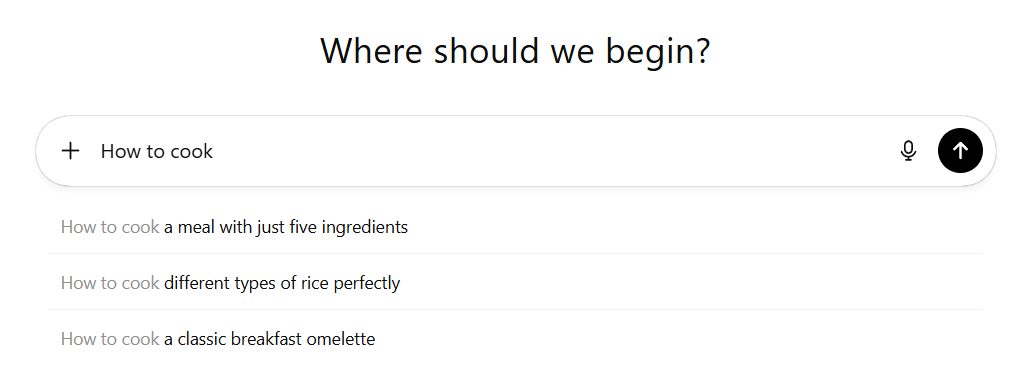}
    \caption{\textcolor{c-color}{Auto-Complete}, ChatGPT}
  \end{subfigure}

  \vspace{0.6em}

  \begin{subfigure}[t]{\linewidth}
    \centering
    \includegraphics[width=1.10\linewidth]{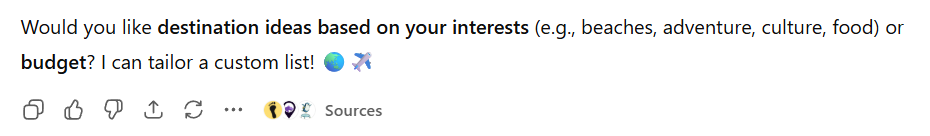}
    \caption{\textcolor{d-color}{Contextual Follow-Ups}, ChatGPT}
  \end{subfigure}
\end{subfigure}
\hfill
% ---------- Right COLUMN ----------
\begin{subfigure}[t]{0.31\textwidth}
  \centering
  % {\large\bfseries PRS \par}
  \vspace{0.1em}
  
  \begin{subfigure}[t]{\linewidth}
    \centering
    \includegraphics[width=0.99\linewidth]{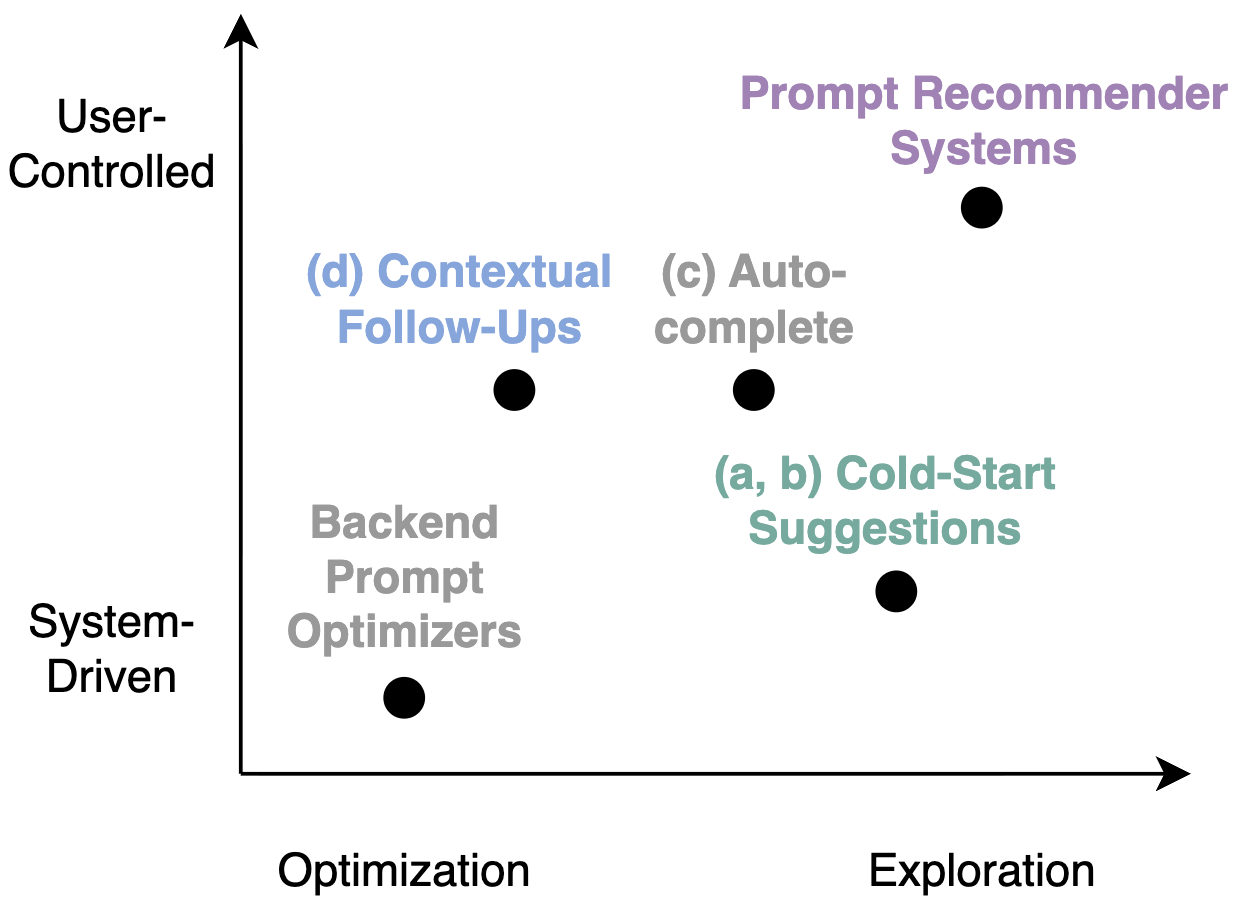}
    \caption{\textbf{Conceptual Overview}}
  \end{subfigure}
\end{subfigure}

\caption{\textbf{We introduce \textcolor{prs-color}{prompt recommender systems} (Figure~\ref{fig:teaser}), a distinct class of prompt interaction methods.} Existing approaches surface recommendations either as \textcolor{a-b-color}{cold-start prompts or static categories at conversation entry (a–b)}, or as \textcolor{c-color}{auto-complete (c)} or \textcolor{d-color}{opportunistic follow-ups appended to model responses (d)}. These designs primarily support prompt initiation or local refinement, but offer limited support for user-controlled exploration, branching, or sustained navigation of prompt space. \textcolor{prs-color}{PRS} instead support exploration as an ongoing interaction, presenting prompts as persistent, selectable alternatives that users can compare and iteratively refine over time (e), treating prompts as recommender-eligible items.}
\label{fig:cold-start}
\end{figure}

\section{Related Work}
\label{sec:related-work}

Prompting is a widely used mode of interaction with large language models (LLMs), in which users express requests through natural language queries that guide model responses \cite{liu2023pre, schulhoff2024prompt, sahoo2024systematic}. Although interaction appears accessible, prior work suggests that prompting is better viewed as a multi-turn, iterative practice rather than a single-shot exchange \cite{nguyen2024beginning, subramonyam2024bridging, 10.1145/3706598.3714259}. Prompt engineering has consequently emerged as a new focus of study \cite{ye2024prompt, sahoo2024systematic, schulhoff2024prompt, marvin2023prompt}.

Prior research identifies several challenges associated with prompting: Users vary widely in their understanding of LLM capabilities and limitations, their ability to articulate effective prompts, and their expectations of system output \cite{nguyen2024beginning, 10.1145/3706598.3714259, KNOTH2024100225}. This disparity conflicts with the apparent openness of the blank chatbox interface, which suggests that any request can be handled effectively~\cite{Koyuturk_2025, sood2025conversationalinterfaceslimitcreativity}. The \textit{gulf of envisioning} describes this gap between users’ intentions and their ability to foresee how prompts shape model behavior, often resulting in limited exploration of the prompt space and poor system outputs \cite{subramonyam2024bridging}. 

In response to these challenges, prior work has explored the distribution, reuse, and management of prompts.
Online communities and marketplaces curate prompt templates for reuse 
(e.g., Reddit \cite{redditPromptGenius}, PromptHub \cite{prompthub}, PromptBase \cite{promptbase}, and Etsy \cite{etsyAIPrompts}), and commercial platforms such as OpenAI provide curated prompt packs for specific job functions and tasks \cite{openaiPromptPacks}.
However, there remains no consensus on what constitutes a high-quality prompt, and real-world prompting often proceeds through trial-and-error \cite{he2025prompting}.
While iterative prompting can improve outputs, it can also be cognitively demanding and may encourage fixation on a narrow range of alternatives, limiting creative exploration \cite{nguyen2024beginning, subramonyam2024bridging}.

These challenges have motivated a wide range of AI support tools aimed at assisting users in constructing, refining, and managing prompts, including templates, guidelines, and automated prompt rewriting or optimization techniques \cite{Khurana_2024, ma2025should, kong2024prewrite, wang2023reprompt, li2024learning}. While effective in some contexts, many such approaches reduce user agency by modifying prompts on users’ behalf or prioritizing optimization over exploration~\cite{10.1145/3706598.3713146}.

\section{Prompt Recommender Systems (PRS)}
\label{sec:prs}

Within this broader landscape, we introduce PRS as a distinct interaction concept for supporting end-user prompting. Traditional recommender systems function as filtering mechanisms that suggest items to drive engagement and support decision-making \cite{raza2026comprehensive}. In contrast, \textit{PRS treat prompts themselves as interactional objects} -- units of intent, strategy, and direction -- that can be surfaced, compared, and iteratively refined during use. Rather than optimizing prompts toward a single ``best'' formulation, PRS aim to expand users’ awareness of possible next steps, supporting exploration within the prompt space.

Prior work related to prompt recommendation remains fragmented and largely task or domain-specific. \citet{tang2025dynamic} explore a dynamic prompt recommender system architecture, integrating domain specific knowledge to improve AI workflows in the cybersecurity domain. Santana et al.~\cite{santana2025responsible, santana2025can} focus on sentence-level prompt recommendations, with the goal of improving prompt quality and adherence to Responsible AI principles. These systems primarily frame prompt recommendation as an object of optimization, rather than as useful for user exploration of the prompt space. Similarly, many widely deployed chatbots -- including ChatGPT and Gemini -- provide prompt recommendations in the form of cold-start suggestions or opportunistic follow-ups at response boundaries (Figure~\ref{fig:cold-start}). These features typically extend autocomplete paradigms to LLM interaction, offering isolated suggestions without supporting comparison, branching, or sustained ideation. As a result, they provide limited assistance for navigating prompt space as an evolving creative process.

Motivated by this gap, we explore and evaluate a PRS that provides in-context follow-up prompts during real writing tasks. We focus on writing because it is a domain with a low barrier to entry and provides rich opportunities for creativity, allowing us to examine how treating prompts as recommender-eligible items shapes exploration, expressiveness, and user experience in generative AI workflows. We examine both academic and creative writing tasks; in general, academic writing expects structure while creative writing expects originality.

\section{\textsc{PromptHelper} User Study}
\label{sec:user-study}

\paragraph{Study Design}
We employed a fully within-subjects 2x2 experimental design crossing
\textit{Task} (Creative vs.\ Academic) with \textsc{PromptHelper} \textit{System State} (ON vs.\ OFF).
Each participant completed all four conditions.
To mitigate order effects, we used a four-group Latin-square counterbalancing scheme, detailed in Appendix \ref{sec:study-design-details}.
The full set of study questions and their mappings to composite measures are provided in Table~\ref{tab:study-questions}.

\paragraph{Participants}
We recruited 32 participants via the Prolific platform,\footnote{\url{https://www.prolific.com/data-annotation}} with equal assignment across the four counterbalancing groups.\footnote{An a priori power analysis assuming a medium effect size ($f = 0.25$), $\alpha = .05$, desired power $= .80$, correlation among measures $= 0.50$, and nonsphericity correction $\epsilon = 0.75$ indicated a required sample size of 29 participants.}
Participants were compensated at a rate of \$12 USD per hour.
The study was administered using an external survey instrument; no personally identifiable or sensitive information was collected.
Sessions lasted approximately 56 minutes on average (SD = 25 minutes).
Participants reported being generally comfortable using chatbots for writing tasks, used such tools some of the time, and were largely familiar with the concept of prompting.
Additional participant details are reported in Appendix~\ref{sec:study-design-details}.

\begin{figure}
    \centering
    \includegraphics[width=0.85\linewidth]{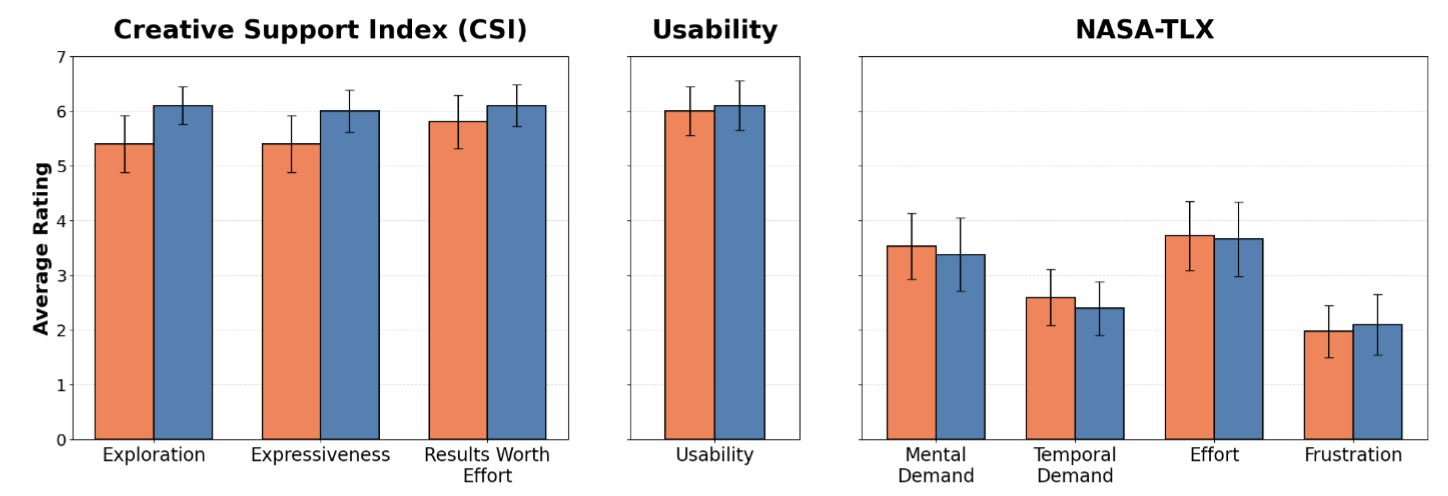}
    \caption{\textbf{Across tasks, \textsc{PromptHelper} improved users’ perceived ability to explore and express ideas, while leaving workload and usability unchanged}, suggesting that prompt recommender systems can scaffold exploratory interaction without imposing additional cognitive burden. \textcolor{figorange}{Orange indicates ratings for the baseline chatbot}, and \textcolor{figblue}{blue indicates ratings for the baseline chatbot with \textsc{PromptHelper} enabled}. We report the mean and standard error of all Likert scales using a 95\% confidence interval. }
    \label{fig:results}
\end{figure}

\paragraph{Tasks}
Before beginning the study tasks, participants were given a brief demonstration of the baseline chatbot, \textit{WritingBot}, implemented using \texttt{gpt-5.1}, as well as the \textsc{PromptHelper} interface.
Each participant completed four 10-minute writing tasks, responding to a provided prompt under each task–system condition.
Following each task, participants completed a post-task survey consisting of four subscales from the NASA Task Load Index (NASA-TLX)~\cite{hart1988development}, three composite measures from the Creativity Support Index (CSI)~\cite{cherry2014quantifying}, and an item assessing perceived usability.
Figure~\ref{fig:results} summarizes the results, and Table~\ref{tab:study-questions} details the mapping between survey items and composite measures.

\section{Results \& Discussion}
\label{sec:results-and-discussion}
We present our findings, organized around key themes that capture how \textsc{PromptHelper} influenced exploration, expressiveness, and workload. We provide the full results in Appendix \ref{sec:extended-results}.

\subsection*{\textbf{$\rhd$ F1: \textsc{PromptHelper} supports exploration of the conversational space.}}
Results from the repeated-measures ANOVA indicate that \textsc{PromptHelper} had a significant main effect on exploration ($p=.001, \eta_p^2=.293$). Follow-up paired-samples t-tests confirmed this effect for both tasks (Academic: $p = .0110$, $d_z = 0.6$; Creative: $p = .0242$, $d_z = .45$). Participants reported that exploration improved in part because the system surfaced alternative directions and tasks the language model could perform, expanding the range of possible next steps during interaction:
\begin{quote}
    \textit{``I found [\textsc{PromptHelper}] had interesting perspectives that went in different directions from my own and made me reconsider a few of my views.''} -- P17, QC4
\end{quote}

\begin{quote}
    \textit{``Prompt Helper is particularly handy in jump starting creativity and overcoming the instances of writers block.''} -- P10, QC7 
\end{quote}

Consistent with these accounts, most participants described the recommended prompts as varied or diverse (QC3). Taken together, these findings suggest that continuous, in-context prompt recommendations can support exploratory interaction over the course of a conversation, rather than serving solely as generic cold-start suggestions.

\subsection*{\textbf{$\rhd$ F2: \textsc{PromptHelper} scaffolds user expressiveness in constrained writing contexts.}}
Results indicate that \textsc{PromptHelper} had a significant main effect on expressiveness ($p=.011, \eta_p^2=.193$). Follow-up analyses show that this effect was significant for academic writing tasks (Academic: $p = .0018$, $d_z = 0.57$), but not for creative writing tasks (Creative: $p = .2043$, $d_z = 0.25$).
These findings suggest that \textsc{PromptHelper} more strongly supports expressiveness in domains with well-established norms and expectations, such as academic writing.

One possible explanation is that academic writing requires adherence to conventions including structure, academic jargon, and factual grounding \cite{itua2014exploring}, which can make it difficult for users to articulate effective prompts without guidance. Prior work has similarly identified challenges in prompting within constrained or technical domains, such as programming \cite{nguyen2024beginning, Khurana_2024} and visual art \cite{oppenlaender2025prompting}, where users may lack the language needed to precisely express intent. In such contexts, \textsc{PromptHelper} may help scaffold expressiveness by providing concrete examples and templates that clarify how requests can be framed:
\begin{quote}
    \textit{``Interacting with Prompt Helper made it easier to decide what to ask next because it reduced the uncertainty around how to phrase a prompt. Seeing examples helped me quickly identify what kind of information the system could use, and that made the next step feel more obvious.``} -- P2, QC6
\end{quote}

In contrast, Results Worth Effort did not show a significant interaction, indicating that participants did not report improved outcomes relative to the effort required. As discussed in the following section, effort levels remained largely unchanged between conditions. Together, these results suggest that \textsc{PromptHelper} shifts how effort is allocated during prompting rather than reducing effort.

\subsection*{\textbf{$\rhd$ F3: \textsc{PromptHelper} shifts cognitive effort from prompt generation to prompt evaluation.}} 
Across all workload and usability measures, we observed no significant differences between conditions with \textsc{PromptHelper} enabled and disabled ($p > .05$). These results suggest that \textsc{PromptHelper} did not reduce overall cognitive demand or alter perceived usability. However, qualitative feedback indicates that the system changed \textit{how effort was allocated} during interaction rather than the \textit{total amount of effort} required.

Participants frequently reported that prompt recommendations reduced decision fatigue by offering clear, actionable options for how to proceed, particularly when they were uncertain about what to ask next. The availability of contextually relevant suggestions helped participants more quickly identify plausible directions for continuation and supported a smoother iteration process:
\begin{quote}
    \textit{``[\textsc{PromptHelper}] made deciding what to ask next easier by offering contextually relevant suggestions that helped to immediately guide and refine the narrative direction. This pre-filtered approach effectively reduced decision fatigue and allowed for a more efficient iteration cycle during the writing task.''} -- P1, QC6
\end{quote}

Participants also noted that \textsc{PromptHelper} lowered the effort required to initiate new prompts by providing concrete starting points. At the same time, most participants reported modifying recommended prompts or using them primarily as inspiration rather than copying them verbatim (QC5), suggesting a shift in effort from prompt generation toward prompt evaluation and refinement.

\section{Conclusion}
In summary, \textsc{PromptHelper} shows that in-context prompt recommendation can increase users’ perceived exploration and expressiveness without adding cognitive burden. By framing prompts as recommender-eligible items, the system helps users navigate prompt space, especially in constrained writing contexts. These results position prompt recommender systems as a lightweight yet effective interaction technique for supporting exploratory human–AI writing workflows.

%%
%% The acknowledgments section is defined using the "acks" environment
%% (and NOT an unnumbered section). This ensures the proper
%% identification of the section in the article metadata, and the
%% consistent spelling of the heading.

\begin{acks}
We thank Anna Seo Gyeong Choi for the feedback on the manuscript.
AI tools were used to assist with portions of the coding and to help organize and refine writing ideas. All outputs were reviewed, verified, and modified by the authors.
\end{acks}

%%
%% The next two lines define the bibliography style to be used, and
%% the bibliography file.
\bibliographystyle{ACM-Reference-Format}
\bibliography{sample-base}

@String{Computing = "Computing" }

@String{Computer = "{IEEE} Computer" }

@String{Academic = "Academic Press" }

@String{Springer = "Springer-Verlag" }

@inproceedings{Khurana_2024, series={IUI ’24},
   title={Why and When LLM-Based Assistants Can Go Wrong: Investigating the Effectiveness of Prompt-Based Interactions for Software Help-Seeking},
   url={http://dx.doi.org/10.1145/3640543.3645200},
   DOI={10.1145/3640543.3645200},
   booktitle={Proceedings of the 29th International Conference on Intelligent User Interfaces},
   publisher={ACM},
   author={Khurana, Anjali and Subramonyam, Hariharan and Chilana, Parmit K},
   year={2024},
   month=mar, pages={288–303},
   collection={IUI ’24} }

@inproceedings{kong2024prewrite,
  title={Prewrite: Prompt rewriting with reinforcement learning},
  author={Kong, Weize and Hombaiah, Spurthi and Zhang, Mingyang and Mei, Qiaozhu and Bendersky, Michael},
  booktitle={Proceedings of the 62nd Annual Meeting of the Association for Computational Linguistics (Volume 2: Short Papers)},
  pages={594--601},
  year={2024}
}

@inproceedings{li2024learning,
  title={Learning to rewrite prompts for personalized text generation},
  author={Li, Cheng and Zhang, Mingyang and Mei, Qiaozhu and Kong, Weize and Bendersky, Michael},
  booktitle={Proceedings of the ACM Web Conference 2024},
  pages={3367--3378},
  year={2024}
}

@inproceedings{wang2023reprompt,
  title={Reprompt: Automatic prompt editing to refine ai-generative art towards precise expressions},
  author={Wang, Yunlong and Shen, Shuyuan and Lim, Brian Y},
  booktitle={Proceedings of the 2023 CHI conference on human factors in computing systems},
  pages={1--29},
  year={2023}
}

@article{oppenlaender2025prompting,
  title={Prompting AI art: An investigation into the creative skill of prompt engineering},
  author={Oppenlaender, Jonas and Linder, Rhema and Silvennoinen, Johanna},
  journal={International journal of human--computer interaction},
  volume={41},
  number={16},
  pages={10207--10229},
  year={2025},
  publisher={Taylor \& Francis}
}

@article{liu2023pre,
  title={Pre-train, prompt, and predict: A systematic survey of prompting methods in natural language processing},
  author={Liu, Pengfei and Yuan, Weizhe and Fu, Jinlan and Jiang, Zhengbao and Hayashi, Hiroaki and Neubig, Graham},
  journal={ACM computing surveys},
  volume={55},
  number={9},
  pages={1--35},
  year={2023},
  publisher={ACM New York, NY}
}

@article{davenport1998fun,
  title={Fun: A condition of creative research},
  author={Davenport, Glorianna and Holmquist, Lars Erik and Thomas, Maureen},
  journal={IEEE MultiMedia},
  volume={5},
  number={03},
  pages={10--15},
  year={1998},
  publisher={IEEE Computer Society}
}

@article{boekhorst2021fun,
  title={Fun, friends, and creativity: A social capital perspective},
  author={Boekhorst, Janet A and Halinski, Michael and Good, Jessica RL},
  journal={The Journal of Creative Behavior},
  volume={55},
  number={4},
  pages={970--983},
  year={2021},
  publisher={Wiley Online Library}
}

@inproceedings{10.1145/3640794.3665881,
author = {Kalirai, Manveer and Kuzminykh, Anastasia},
title = {You Today, Better Tomorrow: Envisioning the Role of Conversation in Recommender Systems of the Future},
year = {2024},
isbn = {9798400705113},
publisher = {Association for Computing Machinery},
address = {New York, NY, USA},
url = {https://doi.org/10.1145/3640794.3665881},
doi = {10.1145/3640794.3665881},
booktitle = {Proceedings of the 6th ACM Conference on Conversational User Interfaces},
articleno = {61},
numpages = {5},
location = {Luxembourg, Luxembourg},
series = {CUI '24}
}

@article{ma2025should,
  title={What should we engineer in prompts? training humans in requirement-driven llm use},
  author={Ma, Qianou and Peng, Weirui and Yang, Chenyang and Shen, Hua and Koedinger, Ken and Wu, Tongshuang},
  journal={ACM Transactions on Computer-Human Interaction},
  volume={32},
  number={4},
  pages={1--27},
  year={2025},
  publisher={ACM New York, NY}
}

@article{cherry2014quantifying,
  title={Quantifying the creativity support of digital tools through the creativity support index},
  author={Cherry, Erin and Latulipe, Celine},
  journal={ACM Transactions on Computer-Human Interaction (TOCHI)},
  volume={21},
  number={4},
  pages={1--25},
  year={2014},
  publisher={ACM New York, NY, USA}
}

@incollection{hart1988development,
  title={Development of NASA-TLX (Task Load Index): Results of empirical and theoretical research},
  author={Hart, Sandra G and Staveland, Lowell E},
  booktitle={Advances in psychology},
  volume={52},
  pages={139--183},
  year={1988},
  publisher={Elsevier}
}

@inproceedings{santana2025responsible,
  title={Responsible Prompting Recommendation: Fostering Responsible AI Practices in Prompting-Time},
  author={Santana, Vagner Figueredo de and Berger, Sara E and Candello, Heloisa and Machado, Tiago and Sanctos, Cassia Sampaio and Su, Tianyu and Williams, Lemara},
  booktitle={Proceedings of the 2025 CHI Conference on Human Factors in Computing Systems},
  pages={1--30},
  year={2025}
}

@article{tang2025dynamic,
  title={Dynamic Context-Aware Prompt Recommendation for Domain-Specific AI Applications},
  author={Tang, Xinye and Zhai, Haijun and Belwal, Chaitanya and Thayanithi, Vineeth and Baumann, Philip and Roy, Yogesh K},
  journal={arXiv preprint arXiv:2506.20815},
  year={2025}
}

@inproceedings{nguyen2024beginning,
  title={How beginning programmers and code llms (mis) read each other},
  author={Nguyen, Sydney and Babe, Hannah McLean and Zi, Yangtian and Guha, Arjun and Anderson, Carolyn Jane and Feldman, Molly Q},
  booktitle={Proceedings of the 2024 CHI Conference on Human Factors in Computing Systems},
  pages={1--26},
  year={2024}
}

@inproceedings{subramonyam2024bridging,
  title={Bridging the gulf of envisioning: Cognitive challenges in prompt based interactions with llms},
  author={Subramonyam, Hari and Pea, Roy and Pondoc, Christopher and Agrawala, Maneesh and Seifert, Colleen},
  booktitle={Proceedings of the 2024 CHI Conference on Human Factors in Computing Systems},
  pages={1--19},
  year={2024}
}

@inproceedings{he2025prompting,
  title={Prompting in the Dark: Assessing Human Performance in Prompt Engineering for Data Labeling When Gold Labels Are Absent},
  author={He, Zeyu and Naphade, Saniya and Huang, Ting-Hao Kenneth},
  booktitle={Proceedings of the 2025 CHI Conference on Human Factors in Computing Systems},
  pages={1--33},
  year={2025}
}

@article{sahoo2024systematic,
  title={A systematic survey of prompt engineering in large language models: Techniques and applications},
  author={Sahoo, Pranab and Singh, Ayush Kumar and Saha, Sriparna and Jain, Vinija and Mondal, Samrat and Chadha, Aman},
  journal={arXiv preprint arXiv:2402.07927},
  year={2024}
}

@inproceedings{marvin2023prompt,
  title={Prompt engineering in large language models},
  author={Marvin, Ggaliwango and Hellen, Nakayiza and Jjingo, Daudi and Nakatumba-Nabende, Joyce},
  booktitle={International conference on data intelligence and cognitive informatics},
  pages={387--402},
  year={2023},
  organization={Springer}
}

@misc{fui2023generative,
  title={Generative AI and ChatGPT: Applications, challenges, and AI-human collaboration},
  author={Fui-Hoon Nah, Fiona and Zheng, Ruilin and Cai, Jingyuan and Siau, Keng and Chen, Langtao},
  journal={Journal of information technology case and application research},
  volume={25},
  number={3},
  pages={277--304},
  year={2023},
  publisher={Taylor \& Francis}
}

@article{wei2022emergent,
  title={Emergent abilities of large language models},
  author={Wei, Jason and Tay, Yi and Bommasani, Rishi and Raffel, Colin and Zoph, Barret and Borgeaud, Sebastian and Yogatama, Dani and Bosma, Maarten and Zhou, Denny and Metzler, Donald and others},
  journal={arXiv preprint arXiv:2206.07682},
  year={2022}
}

@article{schulhoff2024prompt,
  title={The prompt report: a systematic survey of prompt engineering techniques},
  author={Schulhoff, Sander and Ilie, Michael and Balepur, Nishant and Kahadze, Konstantine and Liu, Amanda and Si, Chenglei and Li, Yinheng and Gupta, Aayush and Han, HyoJung and Schulhoff, Sevien and others},
  journal={arXiv preprint arXiv:2406.06608},
  year={2024}
}

@article{raza2026comprehensive,
  title={A comprehensive review of recommender systems: Transitioning from theory to practice},
  author={Raza, Shaina and Rahman, Mizanur and Kamawal, Safiullah and Toroghi, Armin and Raval, Ananya and Navah, Farshad and Kazemeini, Amirmohammad},
  journal={Computer Science Review},
  volume={59},
  pages={100849},
  year={2026},
  publisher={Elsevier}
}

@inproceedings{santana2025can,
  title={Can LLMs Recommend More Responsible Prompts?},
  author={Santana, Vagner Figueredo de and Berger, Sara and Machado, Tiago and de Macedo, Maysa Malfiza Garcia and Sanctos, Cassia Sampaio and Williams, Lemara and Wu, Zhaoqing},
  booktitle={Proceedings of the 30th International Conference on Intelligent User Interfaces},
  pages={298--313},
  year={2025}
}

@article{van2019chatbot,
  title={Chatbot advertising effectiveness: When does the message get through?},
  author={Van den Broeck, Evert and Zarouali, Brahim and Poels, Karolien},
  journal={Computers in Human Behavior},
  volume={98},
  pages={150--157},
  year={2019},
  publisher={Elsevier}
}

@article{itua2014exploring,
  title={Exploring barriers and solutions to academic writing: Perspectives from students, higher education and further education tutors},
  author={Itua, Imose and Coffey, Margaret and Merryweather, David and Norton, Lin and Foxcroft, Angela},
  journal={Journal of further and Higher Education},
  volume={38},
  number={3},
  pages={305--326},
  year={2014},
  publisher={Taylor \& Francis}
}

@misc{redditPromptGenius,
  author       = {{Reddit}},
  title        = {r/ChatGPTPromptGenius},
  howpublished = {\url{https://www.reddit.com/r/ChatGPTPromptGenius/}},
  note         = {Accessed: 2026-01-20},
  year         = {n.d.}
}

@misc{prompthub,
  author       = {{PromptHub}},
  title        = {PromptHub},
  howpublished = {\url{https://prompthub.us}},
  note         = {Accessed: 2026-01-20},
  year         = {n.d.}
}

@misc{promptbase,
  author       = {{PromptBase}},
  title        = {PromptBase},
  howpublished = {\url{https://promptbase.com/}},
  note         = {Accessed: 2026-01-20},
  year         = {n.d.}
}

@misc{etsyAIPrompts,
  author       = {{Etsy}},
  title        = {AI Prompts Marketplace},
  howpublished = {\url{https://www.etsy.com/market/ai_prompts}},
  note         = {Accessed: 2026-01-20},
  year         = {n.d.}
}

@misc{openaiPromptPacks,
  author       = {{OpenAI Academy}},
  title        = {Prompt Packs},
  howpublished = {\url{https://academy.openai.com/public/tags/prompt-packs-6849a0f98c613939acef841c}},
  note         = {Accessed: 2026-01-20},
  year         = {n.d.}
}

@inproceedings{10.1145/3711896.3737856,
author = {Tabari, Narges and Deshmukh, Aniket and Kang, Wang-Cheng and McAuley, Julian and Caverlee, James and Shah, Neil and Karypis, George},
title = {Second Workshop on Generative AI for Recommender Systems and Personalization},
year = {2025},
isbn = {9798400714542},
publisher = {Association for Computing Machinery},
address = {New York, NY, USA},
url = {https://doi.org/10.1145/3711896.3737856},
doi = {10.1145/3711896.3737856},
booktitle = {Proceedings of the 31st ACM SIGKDD Conference on Knowledge Discovery and Data Mining V.2},
pages = {6294–6295},
numpages = {2},
location = {Toronto ON, Canada},
series = {KDD '25}
}

@article{vajjala2024cross,
  title={Cross-domain recommendation meets large language models},
  author={Vajjala, Ajay Krishna and Meher, Dipak and Zhu, Ziwei and Rosenblum, David S},
  journal={arXiv preprint arXiv:2411.19862},
  year={2024}
}

@inproceedings{petruzzelli2024instructing,
  title={Instructing and prompting large language models for explainable cross-domain recommendations},
  author={Petruzzelli, Alessandro and Musto, Cataldo and Laraspata, Lucrezia and Rinaldi, Ivan and de Gemmis, Marco and Lops, Pasquale and Semeraro, Giovanni},
  booktitle={Proceedings of the 18th ACM Conference on Recommender Systems},
  pages={298--308},
  year={2024}
}

@inproceedings{magister2025way,
  title={On the way to llm personalization: Learning to remember user conversations},
  author={Magister, Lucie Charlotte and Metcalf, Katherine and Zhang, Yizhe and Ter Hoeve, Maartje},
  booktitle={Proceedings of the First Workshop on Large Language Model Memorization (L2M2)},
  pages={61--77},
  year={2025}
}

@article{wang2024ai,
  title={Ai persona: Towards life-long personalization of llms},
  author={Wang, Tiannan and Tao, Meiling and Fang, Ruoyu and Wang, Huilin and Wang, Shuai and Jiang, Yuchen Eleanor and Zhou, Wangchunshu},
  journal={arXiv preprint arXiv:2412.13103},
  year={2024}
}

@inproceedings{ning2025user,
  title={User-llm: Efficient llm contextualization with user embeddings},
  author={Ning, Lin and Liu, Luyang and Wu, Jiaxing and Wu, Neo and Berlowitz, Devora and Prakash, Sushant and Green, Bradley and O'Banion, Shawn and Xie, Jun},
  booktitle={Companion Proceedings of the ACM on Web Conference 2025},
  pages={1219--1223},
  year={2025}
}

@inproceedings{park2025leveraging,
  title={Leveraging Multimodal LLM for Inspirational User Interface Search},
  author={Park, Seokhyeon and Song, Yumin and Lee, Soohyun and Kim, Jaeyoung and Seo, Jinwook},
  booktitle={Proceedings of the 2025 CHI Conference on Human Factors in Computing Systems},
  pages={1--22},
  year={2025}
}

@inproceedings{finnie2022robots,
  title={The robots are coming: Exploring the implications of openai codex on introductory programming},
  author={Finnie-Ansley, James and Denny, Paul and Becker, Brett A and Luxton-Reilly, Andrew and Prather, James},
  booktitle={Proceedings of the 24th Australasian computing education conference},
  pages={10--19},
  year={2022}
}

@misc{openai2026codex,
  title        = {Codex},
  author       = {{OpenAI}},
  howpublished = {\url{https://openai.com/codex/}},
  year         = {2026},
  note         = {Accessed: 2026-01-20},
}

@inproceedings{ye2024prompt,
  title={Prompt engineering a prompt engineer},
  author={Ye, Qinyuan and Ahmed, Mohamed and Pryzant, Reid and Khani, Fereshte},
  booktitle={Findings of the Association for Computational Linguistics: ACL 2024},
  pages={355--385},
  year={2024}
}

@article{KNOTH2024100225,
title = {AI literacy and its implications for prompt engineering strategies},
journal = {Computers and Education: Artificial Intelligence},
volume = {6},
pages = {100225},
year = {2024},
issn = {2666-920X},
doi = {https://doi.org/10.1016/j.caeai.2024.100225},
url = {https://www.sciencedirect.com/science/article/pii/S2666920X24000262},
author = {Nils Knoth and Antonia Tolzin and Andreas Janson and Jan Marco Leimeister}
}

@misc{barkley2024investigatingrolepromptingexternal,
      title={Investigating the Role of Prompting and External Tools in Hallucination Rates of Large Language Models}, 
      author={Liam Barkley and Brink van der Merwe},
      year={2024},
      eprint={2410.19385},
      archivePrefix={arXiv},
      primaryClass={cs.CL},
      url={https://arxiv.org/abs/2410.19385}, 
}

@misc{sood2025conversationalinterfaceslimitcreativity,
      title={Do Conversational Interfaces Limit Creativity? Exploring Visual Graph Systems for Creative Writing}, 
      author={Abhinav Sood and Maria Teresa Llano and Jon McCormack},
      year={2025},
      eprint={2507.08260},
      archivePrefix={arXiv},
      primaryClass={cs.HC},
      url={https://arxiv.org/abs/2507.08260}, 
}

@inproceedings{10.1145/3544548.3581388,
author = {Zamfirescu-Pereira, J.D. and Wong, Richmond Y. and Hartmann, Bjoern and Yang, Qian},
title = {Why Johnny Can’t Prompt: How Non-AI Experts Try (and Fail) to Design LLM Prompts},
year = {2023},
isbn = {9781450394215},
publisher = {Association for Computing Machinery},
address = {New York, NY, USA},
url = {https://doi.org/10.1145/3544548.3581388},
doi = {10.1145/3544548.3581388},
booktitle = {Proceedings of the 2023 CHI Conference on Human Factors in Computing Systems},
articleno = {437},
numpages = {21},
location = {Hamburg, Germany},
series = {CHI '23}
}

@inbook{Koyuturk_2025,
   title={Understanding Learner-LLM Chatbot Interactions and the Impact of Prompting Guidelines},
   ISBN={9783031984174},
   ISSN={1611-3349},
   url={http://dx.doi.org/10.1007/978-3-031-98417-4_26},
   DOI={10.1007/978-3-031-98417-4_26},
   booktitle={Artificial Intelligence in Education},
   publisher={Springer Nature Switzerland},
   author={Koyuturk, Cansu and Theophilou, Emily and Patania, Sabrina and Donabauer, Gregor and Martinenghi, Andrea and Antico, Chiara and Telari, Alessia and Testa, Alessia and Buršić, Sathya and Garzotto, Franca and Hernandez-Leo, Davinia and Kruschwitz, Udo and Taibi, Davide and Amenta, Simona and Ruskov, Martin and Ognibene, Dimitri},
   year={2025},
   pages={364–377} }

@inproceedings{10.1145/3706598.3713146,
author = {Qin, Peinuan and Yang, Chi-Lan and Li, Jingshu and Wen, Jing and Lee, Yi-Chieh},
title = {Timing Matters: How Using LLMs at Different Timings Influences Writers' Perceptions and Ideation Outcomes in AI-Assisted Ideation},
year = {2025},
isbn = {9798400713941},
publisher = {Association for Computing Machinery},
address = {New York, NY, USA},
url = {https://doi.org/10.1145/3706598.3713146},
doi = {10.1145/3706598.3713146},
booktitle = {Proceedings of the 2025 CHI Conference on Human Factors in Computing Systems},
articleno = {25},
numpages = {16},
location = {
},
series = {CHI '25}
}

@inproceedings{10.1145/3706598.3714259,
author = {Riche, Nathalie and Offenwanger, Anna and Gmeiner, Frederic and Brown, David and Romat, Hugo and Pahud, Michel and Marquardt, Nicolai and Inkpen, Kori and Hinckley, Ken},
title = {AI-Instruments: Embodying Prompts as Instruments to Abstract \& Reflect Graphical Interface Commands as General-Purpose Tools},
year = {2025},
isbn = {9798400713941},
publisher = {Association for Computing Machinery},
address = {New York, NY, USA},
url = {https://doi.org/10.1145/3706598.3714259},
doi = {10.1145/3706598.3714259},
booktitle = {Proceedings of the 2025 CHI Conference on Human Factors in Computing Systems},
articleno = {1104},
numpages = {18},
location = {
},
series = {CHI '25}
}

%%
%% If your work has an appendix, this is the place to put it.
\appendix

\section{\textsc{PromptHelper} Design Details}
\label{sec:prompthelper-design-details}

\textsc{PromptHelper} defines prompt items using a structured template -- shown in Figure~\ref{fig:prompthelper-item} -- intended to balance conciseness, adaptability, and usability. Prompt items are generated using \texttt{gpt-5.1},\footnote{\url{https://platform.openai.com/docs/models/gpt-5.1}}
 and consist of five fields: \textit{task}, \textit{category}, \textit{context}, \textit{title}, and \textit{prompt}.

\textit{Task} specifies the concrete action the system is proposing (e.g., revising, explaining, or extending prior content). \textit{Category} captures the broader type of support a prompt provides and is selected from a predefined set; categories serve as a primary mechanism for diversifying recommendations and include:
\begin{itemize}
    \item Brainstorming and Ideation - Help generate ideas and develop concepts
    \item Drafting - Write first drafts of various content
    \item Editing and Revision - Refine existing writing by improving clarity, flow, tone, grammar, and structure
    \item Research and fact-checking - Search for current information to support writing
    \item Explanation and Summarization - Explain parts of last answer or summarize content
    \item Structure and organization - Help outline complex pieces, reorganize content for better flow
    \item Feedback - Provide constructive critique on writing, pointing out strengths and areas for improvement.
\end{itemize}

Next, \textit{Context} consists of relevant excerpts from the model’s most recent response, grounding recommendations in the ongoing interaction. \textit{Title} provides a concise, human-readable description of the prompt’s intent and is displayed in the recommendation list. \textit{Prompt} contains the full text of the recommended follow-up and is designed to be directly copied and optionally edited by the user.

During generation, we instructed the LLM to account for multiple plausible user reactions to the prior response, including confusion, dissatisfaction, or interest in further elaboration. Recommended prompts include bracketed alternatives (e.g., [option 1 / option 2 / option 3]) to encourage user modification and preserve flexibility rather than prescribing a single formulation. Prompts are limited to one or two sentences to maintain readability and reduce interaction overhead.

\section{Study Design Details}
\label{sec:study-design-details}
We employed a 2 (Task: \textit{Creative} vs.\ \textit{Academic}) $\times$ 2 (\textsc{PromptHelper}: ON vs.\ OFF) fully within-subjects design. Each participant completed all four task--system combinations. To mitigate order effects, we implemented a four-group Latin-square counterbalancing scheme, with participants randomly and evenly assigned to one of four predetermined condition sequences. Each sequence includes all conditions exactly once while rotating their order across groups.

The four counterbalanced orders were:

\begin{itemize}
    \item \textbf{Group 1}: Creative--ON $\rightarrow$ Creative--OFF $\rightarrow$ Academic--ON $\rightarrow$ Academic--OFF
    \item \textbf{Group 2}: Creative--OFF $\rightarrow$ Academic--ON $\rightarrow$ Academic--OFF $\rightarrow$ Creative--ON
    \item \textbf{Group 3}: Academic--ON $\rightarrow$ Academic--OFF $\rightarrow$ Creative--ON $\rightarrow$ Creative--OFF
    \item \textbf{Group 4}: Academic--OFF $\rightarrow$ Creative--ON $\rightarrow$ Creative--OFF $\rightarrow$ Academic--ON
\end{itemize}

Participants were assigned evenly across groups (8 per group), yielding a total sample size of $N = 32$. We provide the Privacy Policy in Figure~\ref{fig:privacy-policy}.

\begin{figure}[t]
\centering \small
\fbox{%
  \parbox{0.99\linewidth}{%
    {\ttfamily 
\{\\
"task": "Revise analysis",\\
    "category": "Editing and Revision",\\
    "context": "The answer discussed emotional tone but not specific imagery.",\\
    "title": "Revise analysis to include imagery",\\
    "prompt": "Revise your analysis by adding specific imagery to support your discussion of emotional tone."\\
\}
    }
  }%
}
\caption{\textbf{Example prompt item} provided to the LLM, illustrating the structured fields (\textit{task}, \textit{category}, \textit{context}, \textit{title}, and \textit{prompt}) used to define a recommendation.}
\label{fig:prompthelper-item}
\end{figure}

\begin{teaserfigure}[t]
\centering

\begin{subfigure}[t]{0.49\textwidth}
  \centering
  \includegraphics[width=\linewidth]{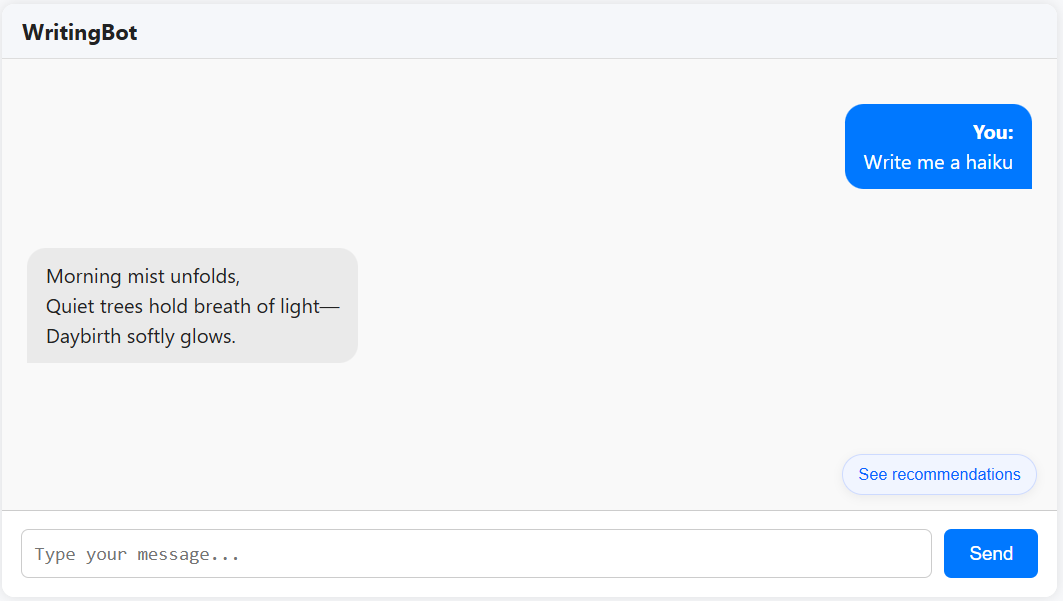}
  \caption{\textsc{PromptHelper} Closed}
\end{subfigure}
\hfill
\begin{subfigure}[t]{0.49\textwidth}
  \centering
  \includegraphics[width=\linewidth]{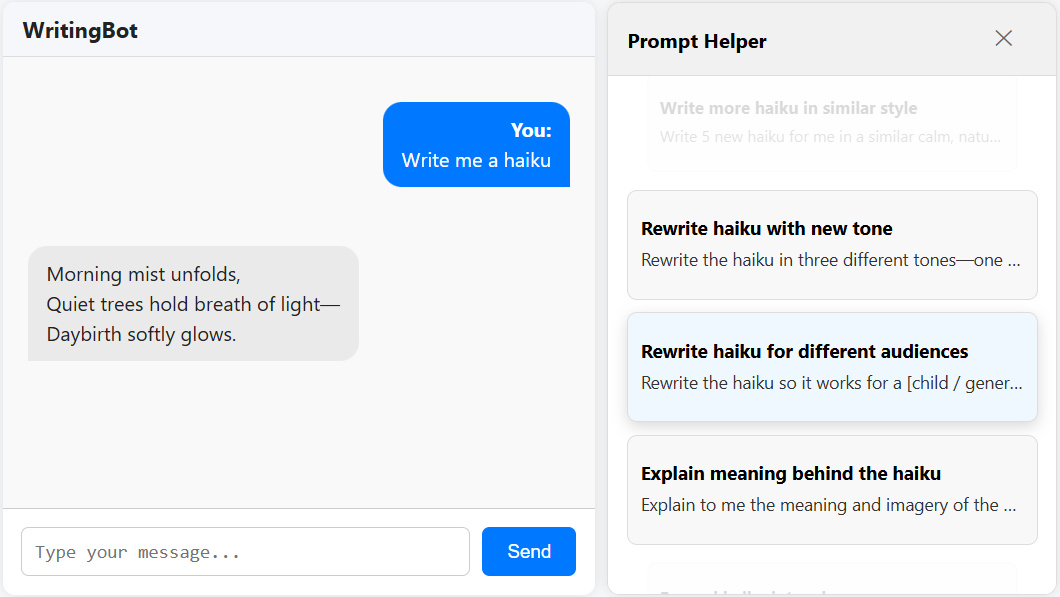}
  \caption{\textsc{PromptHelper} Opened}
\end{subfigure}

\caption{\textsc{PromptHelper} Modes (ON/OFF)}
\end{teaserfigure}

\section{Extended Results}
\label{sec:extended-results}

In this section, we report the extended results. 

\subsection{Repeated-Measures ANOVA Results}
\label{sec:Repeated-Measures ANOVA Results}

This appendix reports the full results of the repeated-measures analyses conducted for the study, with the study design detailed in Appendix~\ref{sec:study-design-details}.

For each dependent measure and composite score, we conducted a repeated-measures ANOVA with 
\textit{Task} and \textit{System State} as within-subjects factors. Because each factor had only two 
levels, sphericity was automatically satisfied and no corrections were required. Degrees of freedom 
were therefore $(1, 31)$ for all main effects and interactions. Partial eta-squared ($\eta_p^2$) is 
reported as the measure of effect size.

The tables below present the sums of squares (SS), mean squares (MS), degrees of freedom (df), 
$F$-statistics, $p$-values, and partial eta-squared values for each effect for the following measures:
Exploration, Expressiveness, Results Worth Effort, Usability, Frustration, and Mental Effort. 
These results provide the full statistical detail underlying the analyses summarized in the main text.

\begin{table}[h!]
\centering
\begin{tabular}{lccccccc}
\toprule
Source & SS & df & MS & $F$ & $p$ & $\eta_p^2$ & Sig. \\
\midrule
\multicolumn{8}{c}{\textbf{Exploration}} \\
\midrule
Task (condition) & 0.5000 & 1,31 & 0.5000 & 0.51 & .479 & .016 & \no \\
System (state) & 14.4453 & 1,31 & 14.4453 & 12.82 & .001 & .293 & \yes \\
Task $\times$ System & 0.3828 & 1,31 & 0.3828 & 0.44 & .512 & .014 & \no \\
\midrule
\multicolumn{8}{c}{\textbf{Expressiveness}} \\
\midrule
Task (condition) & 0.0957 & 1,31 & 0.0957 & 0.14 & .710 & .005 & \no \\
System (state) & 8.7676 & 1,31 & 8.7676 & 7.41 & .011 & .193 & \yes \\
Task $\times$ System & 1.0332 & 1,31 & 1.0332 & 1.95 & .172 & .059 & \no \\
\midrule
\multicolumn{8}{c}{\textbf{Results Worth Effort}} \\
\midrule
Task (condition) 
& 4.1843 & 1,31 & 4.1843 & 5.99 & .020 & .162 & \yes \\
System (state) 
& 3.1698 & 1,31 & 3.1698 & 2.48 & .126 & .074 & \no \\
Task $\times$ System 
& 0.0057 & 1,31 & 0.0057 & 0.01 & .922 & .000 & \no \\
\midrule
\multicolumn{8}{c}{\textbf{Usability}} \\
\midrule
Task (condition) 
& 2.8203 & 1,31 & 2.8203 & 3.58 & .068 & .103 & \no \\
System (state) 
& 0.3828 & 1,31 & 0.3828 & 0.25 & .618 & .008 & \no \\
Task $\times$ System 
& 1.7578 & 1,31 & 1.7578 & 1.91 & .177 & .058 & \no \\
\midrule
\multicolumn{8}{c}{\textbf{Frustration}} \\
\midrule
Task (condition) & 0.0313 & 1,31 & 0.0313 & 0.02 & .882 & .001 & \no \\
System (state) & 0.5000 & 1,31 & 0.5000 & 0.29 & .593 & .009 & \no \\
Task $\times$ System & 0.7813 & 1,31 & 0.7813 & 0.91 & .348 & .028 & \no \\
\midrule
\multicolumn{8}{c}{\textbf{Mental Demand}} \\
\midrule
Task (condition) & 0.5000 & 1,31 & 0.5000 & 0.41 & .525 & .013 & \no \\
System (state) & 0.7813 & 1,31 & 0.7813 & 0.29 & .596 & .009 & \no \\
Task $\times$ System & 0.1250 & 1,31 & 0.1250 & 0.23 & .635 & .007 & \no \\
\midrule
\multicolumn{8}{c}{\textbf{Temporal Demand}} \\
\midrule
Task (condition) & 0.1953 & 1,31 & 0.1953 & 0.14 & .715 & .004 & \no \\
System (state) & 1.3203 & 1,31 & 1.3203 & 1.49 & .231 & .046 & \no \\
Task $\times$ System & 1.3203 & 1,31 & 1.3203 & 1.91 & .177 & .058 & \no \\
\midrule
\multicolumn{8}{c}{\textbf{Effort}} \\
\midrule
Task (condition) & 0.7813 & 1,31 & 0.7813 & 0.34 & .563 & .011 & \no \\
System (state) & 0.1250 & 1,31 & 0.1250 & 0.04 & .841 & .001 & \no \\
Task $\times$ System & 2.5313 & 1,31 & 2.5313 & 1.96 & .171 & .060 & \no \\
\bottomrule
\end{tabular}
\caption{\textbf{Repeated-measures ANOVA results} for within-subjects effects of task (creative vs.\ academic), system state (PromptHelper ON vs.\ OFF), and their interaction across all dependent measures. Reported statistics include sums of squares (SS), degrees of freedom (df), mean squares (MS), $F$-values, $p$-values, and partial eta-squared ($\eta_p^2$). Specific questions detailed in Table~\ref{tab:study-questions}.}
\label{tab:full-anova-results}
\end{table}

\begin{table}[h!]
\centering
\begin{tabular}{lccccc}
\toprule
Comparison & $t$ & $p_{\text{unc}}$ & $p_{\text{corr}}$ & Sig. & $d_z$ \\
\midrule
\multicolumn{6}{c}{\textbf{Exploration}} \\
\midrule
Academic Task: ON vs.\ OFF & 2.98 & .0055 & .0110 & \yes & 0.60 \\
Creative Task: ON vs.\ OFF & 2.37 & .0242 & .0242 & \yes & 0.45 \\
\midrule
\multicolumn{6}{c}{\textbf{Expressiveness}} \\
\midrule
Academic Task: ON vs.\ OFF & 3.67 & .0009 & .0018 & \yes & 0.57 \\
Creative Task: ON vs.\ OFF & 1.30 & .2043 & .2043 & \no  & 0.25 \\
\midrule
\multicolumn{6}{c}{\textbf{Results Worth Effort}} \\
\midrule
Academic Task vs.\ Creative Task in the ON state  & 1.92 & .0636 & .1272 & \no & 0.32 \\
Academic Task vs.\ Creative Task in the OFF state & 1.72 & .0949 & .1272 & \no & 0.27 \\
\bottomrule
\end{tabular}
\caption{\textbf{For significant composites from the ANOVA (shown in Table~\ref{tab:full-anova-results}), we conduct follow-up pairwise $t$-tests with Holm-Bonferroni corrections for multiple comparisons}, and show the results here.}
\end{table}

\subsection{Pairwise $t$-Test Results}
\label{sec:pairwise-t-test}
To further examine the specific effects underlying the repeated-measures ANOVA results, we conducted 
the set of planned paired-samples $t$-tests. These comparisons probed:
\begin{enumerate}
    \item the effect of the \textsc{PromptHelper} \textit{System State} (ON vs. OFF) within each task 
    \item the effect of the \textit{Task} (Academic vs. Creative) within each system state 
\end{enumerate}

Contrasts were tested depending if repeated-measures ANOVA results were significant:
\begin{enumerate}
    \item Academic Task: ON vs.\ OFF 
    \item Creative Task: ON vs.\ OFF 
    \item Academic Task vs.\ Creative Task in the ON state
    \item Academic Task vs.\ Creative Task in the OFF state
\end{enumerate}

All tests used paired-samples $t$-statistics with $df = 31$, and Holm--Bonferroni correction was applied 
to control the familywise error rate across the four comparisons per measure. 
Effect sizes are reported as Cohen's $d_z$ for within-subject contrasts.

The tables below report the $t$-values, uncorrected and corrected $p$-values, significance status, and 
effect sizes for the Exploration and Expressiveness composites.

\begin{table}[h!]
\centering
\small
\begin{tabular}{llp{11cm}}
\toprule
Participant & Question & Quote\\
\midrule
\multicolumn{3}{l}{\cellcolor{lightgray} On \textbf{saving time/efficiency}:}\\
\midrule
P31 & Q6 & \textit{``It made it easy by saving me the time I would have used to come up with the follow-up prompts''}\\
P10 & QC1 & \textit{``The recommendations provided by Prompt Helper were comprehensible and understandable in most cases. The instances of some suggestions being unpredictable or slightly off topic were few, and meant that I had to think hard about whether to implement them or not, but on the whole, these were few and did not interfere with the workflow.''}\\
P2 & Q8 & \textit{``Interacting with Prompt Helper made it easier to decide what to ask next because it reduced the uncertainty around how to phrase a prompt. Seeing examples helped me quickly identify what kind of information the system could use, and that made the next step feel more obvious. It also saved time by giving me a clearer direction instead of having to come up with every prompt from scratch.''}\\
\midrule
\multicolumn{3}{l}{\cellcolor{lightgray} On \textbf{considering new perspectives} and \textbf{navigation of the prompt space}:}\\
\midrule
P29 & Q4 & \textit{``I was able to gain a new perspective that will help my story. I wouldn’t have discovered it on my own if it hadn’t been shown to me.''}\\
P17 & Q4 & \textit{``I let the Prompt Helper design the road map for the conversation. I was curious to see how close it could get to mirroring my thoughts. I found it had interesting perspectives that went in different directions form my own and made me reconsider a few of my views.''}\\
\midrule
\multicolumn{3}{l}{\cellcolor{lightgray} On \textbf{usability}:}\\
\midrule
P29 & Q1 & \textit{``I was fairly comfortable using the writing bot. However, I had some uncertain moments with the prompt helper because I couldn’t figure out how to use it at that time.''}\\
P23 & Q7 & \textit{``Everything worked fine. Just the ``when do I stop'' part was the only confusing thing.''}\\
P7 & Q7 & \textit{``I copied verbatim, I didn't want to confuse the AI with what I was asking.''}\\
\midrule
\multicolumn{3}{l}{\cellcolor{lightgray} On \textbf{fun}, a precondition for creative knowledge work \cite{davenport1998fun, boekhorst2021fun}:} \\
\midrule
P30 & Q7 & \textit{``This was a fun task.''}\\
\bottomrule
\end{tabular}
\caption{\textbf{Representative participant quotes} illustrating perceived efficiency gains, exploration of new perspectives, usability considerations, and enjoyment when using \textsc{PromptHelper}.}
\label{fig:privacy-policy}
\end{table}

\begin{table}[h!]
\centering
\small
\begin{tabular}{llllp{6.7cm}}
\toprule
Q & Type & Source & Composite & Question\\

\midrule
\multicolumn{5}{l}{\cellcolor{lightgray} \textbf{(A) Pre-Survey Questions} -- Performed once before the survey.}\\
\midrule

QA1 & 
\pastelbox{pastelOrange}{Open-Ended} & - & - & 
What experience do you have with using chatbots (e.g. ChatGPT, Gemini, Copilot, Claude, etc.)?\\	

QA2 &
\pastelbox{pastelPink}{Likert} & - & - & 
How often do you use such tools for writing academic or creative pieces?\\

QA3 &
\pastelbox{pastelPink}{Likert} & - & - & 
How comfortable are you with using these tools in writing?\\

QA4 &
\pastelbox{pastelOrange}{Open-Ended} &  - & - & 
What do you know about prompting?\\

\midrule
\multicolumn{5}{l}{\cellcolor{lightgray} \textbf{(B) Survey Questions} -- Performed four times after each 10-minute writing task in the survey.}\\
\midrule

QB1 &
\pastelbox{pastelPink}{Likert} &
\pastelbox{pastelPurple}{CSI} &
\pastelbox{pastelMint}{Exploration} & 
It was easy for me to explore many different ideas, options, designs, or outcomes, using this system.	\\
    
QB2 &
\pastelbox{pastelPink}{Likert} &
\pastelbox{pastelPurple}{CSI} & 
\pastelbox{pastelMint}{Exploration} & 
The system was helpful in allowing me to track different ideas, outcomes, or possibilities.	\\
    
QB3 &
\pastelbox{pastelPink}{Likert} &
\pastelbox{pastelPurple}{CSI} & 
\pastelbox{pastelMint}{Expressiveness} & 
I was able to be very creative while doing the activity inside this system.	\\
    
QB4 &
\pastelbox{pastelPink}{Likert} &
\pastelbox{pastelPurple}{CSI} & 
\pastelbox{pastelMint}{Expressiveness} & 
The system allowed me to be very expressive.	\\
    
QB5 &
\pastelbox{pastelPink}{Likert} &
\pastelbox{pastelPurple}{CSI} & 
\pastelbox{pastelMint}{Results-Worth-Effort} & 
I was satisfied with what I got out of the system.	\\
    
QB6 &
\pastelbox{pastelPink}{Likert} &
\pastelbox{pastelPurple}{CSI} & 
\pastelbox{pastelMint}{Results-Worth-Effort} & 
What I was able to produce was worth the effort I had to exert to produce it.	\\
    
QB7 &
\pastelbox{pastelPink}{Likert} &
\pastelbox{pastelBlue}{NASA-TLX} & 
\pastelbox{pastelMint}{Mental Demand} & 
How mentally demanding was the task?	\\
    
QB8 &
\pastelbox{pastelPink}{Likert} &
\pastelbox{pastelBlue}{NASA-TLX} & 
\pastelbox{pastelMint}{Results-Worth-Effort} & 
How hurried or rushed was the pace of the task?	\\

QB9 &
\pastelbox{pastelPink}{Likert} &
\pastelbox{pastelBlue}{NASA-TLX} & 
\pastelbox{pastelMint}{Mental Demand} & 
How hard did you have to work to accomplish your level of performance?	\\
    
QB10 &
\pastelbox{pastelPink}{Likert} &
\pastelbox{pastelBlue}{NASA-TLX} & 
\pastelbox{pastelMint}{Usability} & 
How insecure, discouraged, irritated, stressed, and annoyed were you?	\\
    
QB11 &
\pastelbox{pastelPink}{Likert} & & & 
Self-rate your written response on an (A-F) grading scale.	\\
    
QB12 &
\pastelbox{pastelPink}{Likert} &
\pastelbox{pastelBlue}{NASA-TLX} & 
\pastelbox{pastelMint}{Usability} & 
Rate the overall usability of the system.\\

\midrule
\multicolumn{5}{l}{\cellcolor{lightgray} \textbf{(C) Post-Survey Questions} -- Performed once after the survey.}\\
\midrule

QC1 &
\pastelbox{pastelOrange}{Open-Ended} &  - & - & 
How comfortable were you with using Prompt Helper and WritingBot? Were there any moments where you felt uncertain about what it was doing or suggesting?	\\

QC2 &
\pastelbox{pastelOrange}{Open-Ended} &  - & - & 
How did your approach to writing using WritingBot compare between tasks with and without Prompt Helper?\\

QC3 &
\pastelbox{pastelOrange}{Open-Ended} &  - & - & 
How would you describe the range or variety of prompts suggested by Prompt Helper (e.g., varied, repetitive, surprising, helpful)? 	\\

QC4  &
\pastelbox{pastelOrange}{Open-Ended} &  - & - & 
In what ways did Prompt Helper affect how you generated or developed your ideas? Did it lead you to explore directions you wouldn’t have taken on your own?  	\\

QC5 &
\pastelbox{pastelOrange}{Open-Ended} &  - & - & 
How did you use the prompts being recommended by Prompt Helper (e.g.  ignored, inspired, modified, or copied verbatim)?	\\

QC6 &
\pastelbox{pastelOrange}{Open-Ended} &  - & - & 
Did interacting with Prompt Helper make it easier or harder to decide what to ask next? Please describe why.	\\

QC7 &
\pastelbox{pastelOrange}{Open-Ended} &  - & - & 
Is there anything else you’d like to share about your experience with WritingBot and/or Prompt Helper? \\

\bottomrule
\end{tabular}
\caption{\textbf{Study questionnaire design}: Likert (1--7) and open-ended items mapped to CSI \cite{cherry2014quantifying} and NASA-TLX NASA-TLX \cite{hart1988development} constructs and aggregated into composite measures for analysis.}
\label{tab:study-questions}
\end{table}

\section{Limitations \& Future Directions}
\label{sec:limitations-future-directions}

\paragraph{Supporting Longer Tasks}
Although our study used a controlled, Latin-square within-subjects design, each writing task was limited to 10 minutes which may not be long enough to completely reflect the challenges and complexities faced during longer writing workflows (e.g., book writing). For example, recommendations may become more generic over a longer duration, especially through learning effects. 

\paragraph{Recommendation Intelligence}
\textsc{PromptHelper} intentionally does not incorporate ranking, personalization, or adaptive learning~\cite{10.1145/3711896.3737856, vajjala2024cross, petruzzelli2024instructing, magister2025way, wang2024ai, ning2025user}, enabling us to study prompt recommendation as a lightweight, minimally intrusive interaction technique, without confounding effects from preference learning or automated selection. Future work could explore ranking or personalization, particularly in longer-term workflows, while examining how increased system intelligence shapes trust, agency, and exploratory behavior~\cite{subramonyam2024bridging}. Further, 
\textbf{by releasing the PromptHelper dataset and task formulation, we further position prompt recommendation as a standalone recommendation problem and invite future work to develop and benchmark ranking, personalization, and adaptive methods in this setting.} We release the dataset and code at \url{https://anonymous.4open.science/r/Prompt_Recommender-6B80}. Such work can examine how increased recommendation intelligence shapes trust, agency, and exploratory behavior, particularly in longer-term workflows~\cite{subramonyam2024bridging}.

\paragraph{Interfaces}
We are interested in investigating when and where PRS interfaces are most effective. In particular, future work can examine how similar lightweight, on-demand designs transfer to other AI contexts~\cite{10.1145/3544548.3581388}, including IDEs~\cite{finnie2022robots, openai2026codex} and multimodal authoring environments~\cite{park2025leveraging}, where users already expect structured system support and may benefit from richer forms of prompt scaffolding.

\paragraph{Commercial Content and Advertising}
\textbf{Prompt recommender systems also present a concrete opportunity for integrating commercial content into conversational systems, such as embedding advertisements within chatbots via PRS~\cite{van2019chatbot}.} Because PRS surface follow-up prompts that are contextually relevant, optional, and user-controlled, they offer an LLM-era alternative to traditional advertising techniques. Future work can investigate how PRS-based advertising interfaces shape user trust, agency, and engagement, and whether prompt-level recommendations can support commercial goals while maintaining transparency and alignment with users’ ongoing tasks.

\paragraph{Increasing AI Literacy} 
Prompt recommender systems serve as an opportunity to promote AI literacy. \citet{santana2025responsible} notes that even professionals with technical backgrounds lack the time to study the latest prompting practices. \citet{barkley2024investigatingrolepromptingexternal} shows that poor prompts lead to increased hallucinations -- and therefore worse quality -- in AI outputs. Additionally, \citet{10.1145/3544548.3581388} identifies both a capability gap and tendency for non-AI experts to overgeneralize interactions with an LLM. Prompt recommendation may be able to educate users by providing general examples of wording a prompt and demonstrating different capabilities. Although this work did not directly measure learning outcomes, future research could examine the educational impact of such systems on users’ understanding of AI.

\begin{figure}[t]
\centering
\fbox{%
  \parbox{0.85\linewidth}{%
    {
    \textbf{Privacy Policy}\\
    \textbf{Last Updated:} November 24, 2025 \\
    \textbf{Platform:} Prolific (\url{https://www.prolific.com})\\
    \textbf{1. Purpose of the Task}\\
    We are studying how users utilize a new technology, called prompt recommender systems, in their interactions with Generative AI. Please complete the writing tasks in any way you feel comfortable with. There are no right or wrong ways to use the system — we are interested in how you approach writing with these tools.\\
    \textbf{2. Data We Collect}\\
    During your participation, we collect the following categories of data:
    \begin{itemize}
        \item \textbf{Prolific ID and Prolific-provided Demographic Data:} This data is provided by the Prolific platform to us.
        \item \textbf{Timing Metadata:} The start and completion time for your task submission, which helps assess annotation duration and data quality.
        \item \textbf{Survey Data:} Your survey responses.
        \item \textbf{Log Data:} Your interactions, prompts, all inputs and timing to the software will be stored and logged.
    \end{itemize}
    No personal device data is collected by the researchers, only interaction data with the provided software system and such data collected and distributed by Prolific; such information remains with Prolific and is governed by their Privacy Policy.\\
    \textbf{3. How Your Data Is Used}\\
    Your responses and interactions will be de-identified – used for academic research and publication in peer-reviewed journals or conference presentations. De-identified datasets will be shared publicly for use by other researchers under ethical data-sharing agreements, consistent with open science practices.\\
    \textbf{4. Data Storage and Security}
    \begin{itemize}
    \item All data collected will be stored securely in encrypted storage (e.g., Google cloud) accessible only to the research team.
    \item Data will be retained for up to 3 years after study completion and then deleted or permanently anonymized.
   \item No data will be sold or shared with commercial entities.
    \end{itemize}
    \textbf{5. Voluntary Participation and Withdrawal}\\
    Your participation is entirely voluntary. You may withdraw from the task at any time prior to submission on Prolific. If you withdraw before completing the task, no partial data will be used.\\
    \textbf{6. Risks and Benefits}\\
    There are minimal risks associated with participation. You may experience minor fatigue from reviewing university materials. Your work will contribute to understanding and improving Generative AI systems.\\
    \textbf{7. Confidentiality}\\
    Your responses will never be linked to your name or contact information. Any publications or presentations resulting from this research will contain only anonymous findings (including, e.g., ``P1 said xyz.'').\\
    \textbf{8. Consent}\\
    By completing the task on Prolific, you confirm that you:
    \begin{itemize}
        \item Are 18 years of age or older.
        \item Understand the nature and purpose of this research.
        \item Consent to your anonymized responses being used for research and publication purposes.
    \end{itemize}
    }
  }%
}
\caption{\textbf{Privacy Policy Provided to Human Participants Recruited via the Prolific Platform.}}
\label{fig:privacy-policy}
\end{figure}

\end{document}